\documentclass{article}
\usepackage{spconf,amsmath,graphicx}
\usepackage{mathrsfs}
\usepackage{amsfonts}
\usepackage{color, soul}
\usepackage{multirow}
\usepackage{xspace}
\usepackage{ctable}
\usepackage{makecell}
\usepackage{floatrow}
\usepackage{sidecap}
\usepackage{hyperref}

\usepackage[fontsize=9.7]{scrextend}

\makeatletter
\DeclareRobustCommand\onedot{\futurelet\@let@token\@onedot}
\def\@onedot{\ifx\@let@token.\else.\null\fi\xspace}

\def\eg{\emph{e.g}\onedot\ }

\def\etc{\emph{etc}\onedot} 
\def\wrt{w.r.t\onedot} 
\def\etal{\emph{et al}\onedot}
\makeatother

\newcommand\norm[1]{\left\lVert#1\right\rVert}

\let\oldparagraph=\paragraph
\renewcommand\paragraph[1]{\oldparagraph{#1.}}

\title{Speaker diarization with session-level speaker embedding refinement using graph neural networks}
\name{Jixuan Wang$^{1,2}$\sthanks{Initial work by first author done as an intern at Microsoft.}, Xiong Xiao$^{3}$, Jian Wu$^{3}$, Ranjani Ramamurthy$^{3}$, Frank Rudzicz$^{1,2}$, Michael Brudno$^{1,2}$}
\address{$^{1}$University of Toronto, Canada ~ $^{2}$Vector Institute, Canada ~ $^{3}$Microsoft, USA \\
$^{1,2}$\{jixuan, frank, brudno\}@cs.toronto.edu ~ $^{3}$\{xioxiao, jianwu, ranjanir\}@microsoft.com}
\begin{document}
%
\maketitle

\begin{abstract}
Deep speaker embedding models have been commonly used as a building block for speaker diarization systems; however, the speaker embedding model is usually trained according to a global loss defined on the  training data, which could be sub-optimal for distinguishing speakers locally in a specific meeting session. In this work we present the first use of graph neural networks (GNNs) for the speaker diarization problem, utilizing a GNN to refine speaker embeddings locally using the structural information between speech segments inside each session. The speaker embeddings extracted by a pre-trained model are remapped into a new embedding space, in which the different speakers within a single session are better separated. The model is trained for linkage prediction in a supervised manner by minimizing the difference between the affinity matrix constructed by the refined embeddings and the ground-truth adjacency matrix. Spectral clustering is then applied on top of the refined embeddings. We show that the clustering performance of the refined speaker embeddings outperforms the original embeddings significantly on both simulated and real meeting data, and our system achieves the state-of-the-art result on the NIST SRE 2000 CALLHOME database.
\end{abstract}
\section{Introduction}
\label{sec:intro}
Speaker diarization is the problem of ``who spoke when".
A typical speaker diarization system usually contains multiple steps. 
First, the non-speech parts are filtered out by voice activity detection (VAD). Second, the speech parts are split into small homogeneous segments either uniformly or according to the detected speaker change points.  
Third, each segment is mapped into a fixed dimensional embedding, such as i-vector~\cite{dehak2010front}, x-vector~\cite{snyder2018x} or d-vector~\cite{variani2014deep, bredin2017tristounet, wan2018generalized, wang2019centroid,zhou2019cnn}. Finally clustering methods or end-to-end approaches are applied to generate the diarization results~\cite{meignier2010lium, shum2013unsupervised, sell2014speaker, garcia2017speaker, wang2018speaker, zhang2019fully}. Usually a classifier needs to be trained for similarity scoring for i-vectors and x-vectors, while similarities between d-vectors can usually be measured by simple distance metrics, \eg{cosine or Euclidean distance}.  

Commonly-used clustering methods for speaker diarization include K-means~\cite{lloyd1982least}, agglomerative hierarchical clustering (AHC)~\cite{rokach2005clustering}, spectral clustering (SC)~\cite{von2007tutorial} and affinity propagation~\cite{frey2007clustering}.
Although deep learning methods driven by large scale datasets have dominated the fields of speaker and speech recognition, it is still non-trivial to design an end-to-end objective function for the speaker diarization problems which is permutation invariant in terms of both the speaker order and speaker number.
Most recently there have been several end-to-end approaches that either utilize a factored generative model~\cite{zhang2019fully} or are trained according to a permutation-free loss~\cite{fujita2019end}.

\begin{figure}[tb]
  \centering
  \centerline{\includegraphics[width=9cm]{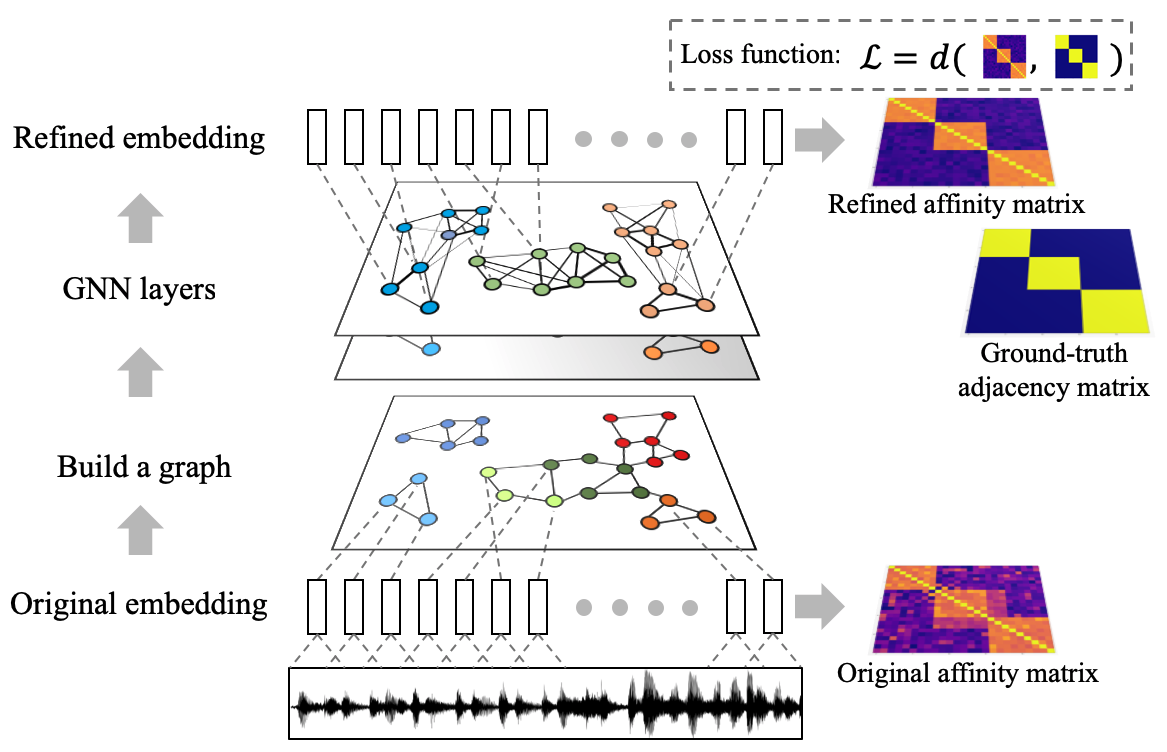}}
\caption{Overview of the proposed method. A graph is constructed for each session using speaker embeddings extracted by a pre-trained speaker embedding model. A GNN model is applied to remap the original embedding into another embedding space  which is trained according to a loss function defined by the difference between the refined affinity matrix and the ground-truth adjacency matrix. }
\label{fig:overview}
\end{figure}

In this paper, we consider a task upstream of speaker diarization.
We suggest that the speaker diarization models could be further improved by locally refining the speaker embedding for each session. The speaker embedding models are usually designed to generalize well across a large range of speakers with different characteristics. However, speaker diarization is a simpler task compared with speaker recognition in the perspective of distinguishing speakers: we only care about separating several speakers for each session.

Our approach is to refine speaker embeddings by improving their performance for similarity measurement between speech segments.
Obviously, if we could correctly predict for each pair of speech segments whether they belong to the same speaker or not, the diarization problem would be solved. 
Our method utilizes local structural information existing among the speaker embedding spaces for each session.
The structural information is expressed by a graph built upon the speech segments for each session. 
The task of predicting similarities between pairs of nodes is analogous to the link prediction task on graphs.
Our model can be seen as a neural link predictor~\cite{kipf2016variational, zhang2018link, nickel2015review} consisting of an \textit{encoding component} and a \textit{scoring component}.
The \textit{encoding component} includes GNN~\cite{wu2019comprehensive, cai2018comprehensive, kipf2016semi, gilmer2017neural} layers to map speaker embeddings into another embedding space. Fully connected layers or cosine similarity is used as the \textit{scoring} component.
To train the model, a loss function is defined as the difference between the affinity matrix constructed by the refined embeddings and the ground-truth adjacency matrix. By minimizing this loss the model will be able to refine the original embeddings so that different speakers could be better separated within each session.

Very recently (and in parallel to this work), Lin \etal{~\cite{lin2019lstm}} proposed the use of LSTMs to improve similarity measurement for speaker diarization. 
Unlike their work, we utilize the structural information in the embedding space instead of utilizing the temporal information in label sequences. 
Also our model not only outputs better similarity measurement but also refined speaker embeddings, which could be potentially fed into non-clustering-based methods. Our model also achieves better performance with the same experimental setup.
Experiments on both simulated and real meeting data show that the performance of speaker number detection and speaker diarization with the refined embeddings outperforms the original embeddings significantly, and our system achieves the state-of-the-art result on the NIST SRE 2000 CALLHOME database.


\section{Graph based speaker diarization}
\label{sec:graph}
\subsection{Building graphs of speech segments}
A graph is built for each session using the pretrained speaker embeddings. Each node represents an audio segment which could be word-level, utterance-level, or extracted by a sliding window with a fixed sliding step. Speaker embedding of each segment is extracted as the node features. The weight of edges between nodes is represented by the PLDA scores or cosine similarities between the corresponding x-vectors and d-vectors, respectively. We only keep edges of weight larger than a threshold, which is treated as a hyperparameter.


Formally, each meeting session can be represented as a graph $\mathcal{G}(\mathcal{V}, \mathcal{E}, A)$, where $\mathcal{V}$ is the set of nodes (speech segments), $\mathcal{E}$ is the set of edges, and $A \in N \times N$ is the affinity matrix with $A_{ij} > 0$ if edge $e_{ij} = (v_i, v_j) \in \mathcal{E}$ and  $A_{ij} = 0$ otherwise.
The matrix of d-vectors $X \in \mathbb{R}^{N \times D}$ can be treated as node features for graph $\mathcal{G}$, where $N$ is the total number of segments and $D$ is the dimension of the embedding space.
The features of each node $v_i$ are represented by a d-vector $\mathbf{x}_i$, $i \in {1, 2, \dots, N}$. The goal of speaker diarization can be formulated as the prediction of $y_i$ for each segment embedding $\mathbf{x}_i$ such that $y_i$ is equal to $y_j$ if and only if $\mathbf{x}_i$ and $\mathbf{x}_j$ belong to the same speakers, $i, j \in {\{1, 2, \dots, N\}}$.

\subsection{Graph neural networks}
We apply several variants of GNNs falling into the framework of message passing neural networks (MPNNs)~\cite{gilmer2017neural}. Under the MPNN framework, the convolutional operator is expressed as a message passing scheme:

\begin{equation}
    \mathbf{x}_i^{\prime} = \gamma_{\mathbf{\Theta}} \left( \mathbf{x}_i, \square_{j \in \mathcal{N}(i)} \, \phi_{\mathbf{\Theta}}\left(\mathbf{x}_i, \mathbf{x}_j,\mathbf{e}_{i,j}\right) \right)
\end{equation}
where $\mathbf{x}_i$ is the feature of node $i$ in the current layer with dimension $D$, $\mathbf{x}_i^{\prime}$ is the node feature in the next layer with dimension $D^{\prime}$, $\mathbf{e}_{i,j}$ is the edge feature from node $i$ to node $j$, $\gamma(\cdot)$ and $\phi(\cdot)$ are the update function and message function, respectively, parameterized by $\mathbf{\Theta}$, and $\square$ denotes the aggregation function, \emph{e.g}\onedot, $sum$, $mean$, $max$, \etc.

One GNN variant we apply is the graph convolutional network (GCN) described by~\cite{kipf2016semi} in which the $\square$ function corresponds to taking a certain weighted average of neighboring nodes. The updating scheme between two layers is: 
\begin{equation}
    \mathbf{x}_i^{\prime} = \sigma \left(W \sum_j L_{ij} \mathbf{x}_j \right),
\end{equation}
where $L = \hat{D}^{-1/2}\hat{A}\hat{D}^{-1/2}$ is the normalized affinity matrix added by self connection, $\hat{A}=A+I_N$, $\hat{D}$ denotes the degree matrix of $\hat{A}$, $W \in \mathbb{R}^{D^{\prime} \times D}$ is a layer-specific trainable weight matrix, and $\sigma(\cdot)$ is a nonlinear function.


\subsection{Model design}
Our model consists of two components: an \textit{encoding component} and a \textit{scoring component}.
GNN layers are applied as the \textit{encoding component}. No nonlinear functions are applied between the GNN layers. Because the speaker embeddings are already well trained, adding nonlinearity may lose information inside the original embeddings and result in worse results.
The \textit{encoding component} generates the refined embeddings, of which every pair are concatenated as inputs to the \textit{scoring component}.

Due to the distinct nature of x-vectors and d-vectors, different \textit{scoring components} are designed.
For d-vectors, it is common to leave the complexity to the model and use a simple distance metric, \eg{cosine distance}. Since the original d-vectors are trained for being comparable by a simple distance metric, we keep using this distance metric for scoring the refined embeddings.
For x-vectors, on the other hand, an additional classifier, \eg{PLDA}, is usually trained to measure the similarity. In our model, we add fully connected layers with nonlinear functions after the GNN layers, which is actually a classifier on top of the refined embeddings.
This classifier is trained together with the GNN layers. 

\subsection{Loss function}
We train the GNN model \wrt a loss function that is defined as the difference between the affinity matrix constructed by the refined embedding and the ground-truth adjacency matrix.
This is actually a binary classification of linkage for all pairs of segments inside each session.
For the x-vector based system, we simply applied the binary cross entropy (BCE) loss on all pairs of segments for each session.

However, this simple loss function does not work for d-vectors. 
Due to the simplicity of the distance metric used, the gradient would be very noisy  if training the model \wrt the exact match between the cosine similarities and ground-truth $1$s or $0$s.
Instead we applied the following loss function:
\begin{equation}
    \mathcal{L} = hist\_loss(A, A_{gt}) + \alpha \cdot \norm{A - A_{gt}}_{nuclear},
    \label{eq:loss}
\end{equation}
in which $hist\_loss$ corresponds to the histogram loss~\cite{ustinova2016learning}, $\norm{\cdot}_{nuclear}$ is the nuclear norm between $A$ and the ground-truth adjacency matrix $A_{gt}$, and $\alpha$ is a scalar value hyperparameter.
Histogram loss tries to ensure the distance between pairs of segments belonging to different speakers is larger than the distance between those belonging to the same speakers. 
The nuclear norm, which is the sum of singular values of a matrix~\cite{recht2010guaranteed} depends on the rank match between $A$ and $A_{gt}$, instead of an exact match. The combination of these two loss functions achieves similarity scores that stay in a reasonable range.

\subsection{Spectral clustering}
We use spectral clustering as our backend method and follow the standard steps described in~\cite{von2007tutorial}. However, to detect the speaker number, we first perform eigen-decomposition on the refined affinity matrix and then determine the speaker number by the number of eigenvalues that are larger than a threshold, which we treat as a tunable parameter. As shown in Section~\ref{sec:spk_num}, this method outperforms the commonly used method based on maximal eigengap.

\section{Experiments with x-vectors}
\subsection{Datasets and implementation details}
To compare our x-vector-based systems with other methods, we follow the evaluation steps described in the \textit{callhome\_diarization/v2} receipe of Kaldi~\cite{callhome_receipe}.
We used the pretrained x-vector model and PLDA scoring model trained on augmented Switchboard and NIST SRE datasets~\cite{xvector_model}.
Results on the commonly used dataset NIST SRE 2000 CALLHOME (LDC2001S97) Disk 8 are reported.

Our model architecture includes two GCN~\cite{kipf2016semi} layers followed by two fully connected layers. Input x-vector is 128-dimensional. The dimension of output embedding from the GCN layers is 64, of which every pair is concatenated as a 128-dimensional input to the following fully connected layers.
The first fully connected layers is 64-dimensional with the ELU activation function~\cite{clevert2015fast} and the last layer has 1 dimension followed by a sigmoid function. The GCN layers are implemented with the PyTorch Geometric library~\cite{fey2019fast}.

Similar to previous work~\cite{zhang2019fully, lin2019lstm}, we conducted 5-fold cross validation. The dataset contains 500 sessions in total, which is uniformly split into 5 subsets of 100 sessions for evaluation. In each turn, the model is trained for 50 epochs on 400 sessions with an initial learning rate of 0.001 reduced by a factor of 10 after 40 epochs. 
The optimal threshold for speaker number detection on the training set is used for evaluation on the testing set. A graph is built for each session on which we connect a pair of nodes if their PLDA score is higher than 0.2. The model is trained in a session-by-session fashion that each batch contains a single graph constructed from one session. All the hyperparameters are tuned on a validation set split from the training set for each turn. Results are shown in Table~\ref{tbl:callhome} and summarized in the following section.
\begin{table}[h]
\caption{DER (\%) on the NIST SRE 2000 CALLHOME dataset. SC refers to spectral clustering and AHC to agglomerative hierarchical clustering.}
\label{tbl:callhome}
\centering
\begin{tabular}{llr}
\specialrule{.1em}{.05em}{.05em} 
& Method                & DER(\%) \\ \hline
\multirow{2}{*}{Baseline} & x-vector + PLDA + AHC \textit{(5-fold)} & 8.64   \\
         & x-vector + PLDA + SC \textit{(5-fold)} & 8.05   \\ \hline
\multirow{5}{*}{\thead{Recent \\ Work}} & Wang \etal{~\cite{wang2018speaker}}           & 12.0   \\
 & Sell \etal{~\cite{7178881}}           & 11.5   \\
 & Romero \etal{~\cite{7953094}}         & 9.9    \\
 & Zhang \etal{~\cite{zhang2019fully}} \textit{(5-fold)} & 8.5    \\
 & Lin \etal{~\cite{lin2019lstm}} \textit{(5-fold)}           & 7.73   \\
 \hline
Ours & GNN based \textit{(5-fold)}          & \textbf{7.24}   \\ 
\specialrule{.1em}{.05em}{.05em} 
\end{tabular}
\end{table}

\subsection{Speaker diarization results}
We compare speaker diarization error rate (DER) of our model with several recent works. We achieve state-of-the-art performance resulting in a DER of 7.24\%, which outperforms both the baselines and all recent works. For fairness, we only include the results without system fusion of~\cite{lin2019lstm}.
We only include the result with in-domain training of~\cite{zhang2019fully}. The DER of~\cite{zhang2019fully} with external data is 7.6\% which is still worse than our result. Also the speaker embedding model of~\cite{zhang2019fully} was trained on a much larger dataset than the one we are using, which may also influence the results.    
\section{Experiments with d-vectors}
\subsection{Datasets and implementation details}
For d-vector-based system, we trained the model on simulated conversations and evaluated it on real meetings. The d-vector extraction model is trained and fake meetings are simulated from the VoxCeleb2 dataset~\cite{chung2018voxceleb2}. For each simulated meeting, we randomly select 2-15 speakers from the training set of VoxCeleb2 as the meeting participants. And for each speaker, 2-60 speech segments with duration of 1.5s are sampled. Moreover, we use 45 sessions of in-house real meetings as another testing set.
On average, each meeting contains 6 speakers, ranging from 2 to 16. The average duration of meeting sessions is about 30 minutes.  

To build the graphs, we only connect a pair of speech segments if the cosine similarity between their d-vectors is larger than a threshold 0.2, which is tuned on the validation set.  
Our model contains two GCN layers, of which the hidden dimension and output dimension are both equal to the input dimension of 128. We used 150 bins for the histogram loss. The DER reported only includes speaker confusion.

\subsection{Speaker number detection results}
\label{sec:spk_num}
\begin{figure}[h]
  \centering
\includegraphics[width=7.0cm]{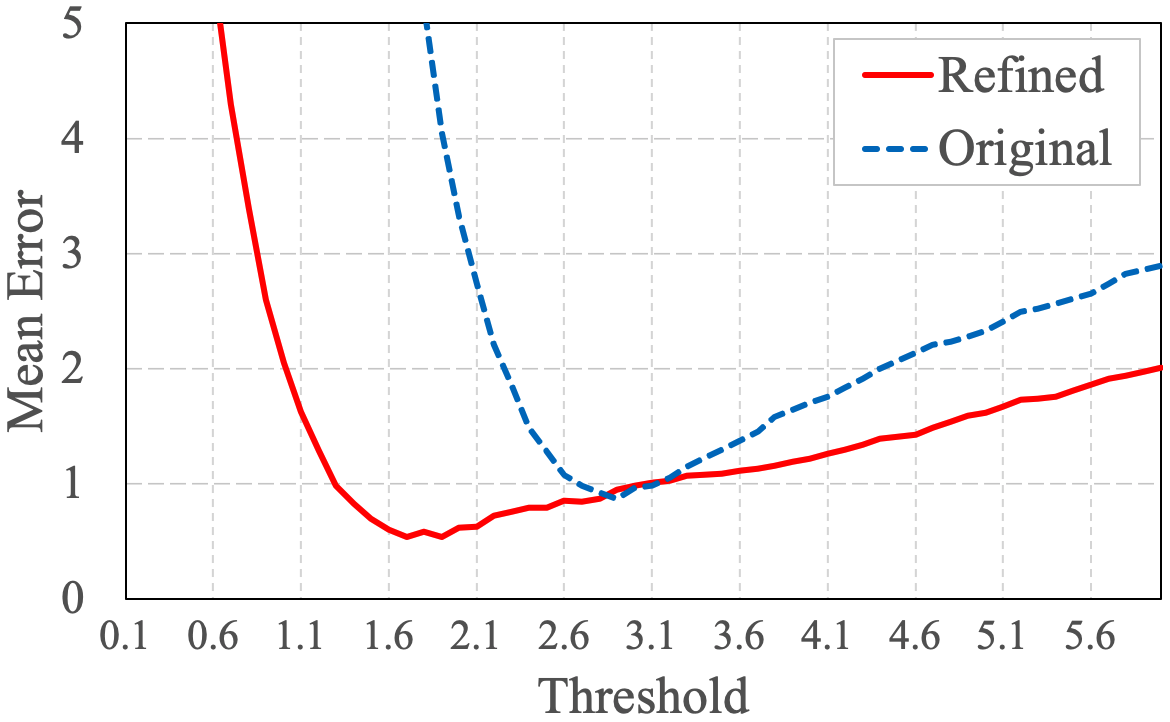}
\caption{Mean error of speaker number detection by different thresholds. The dashed line (original) refers to results with the original speaker embeddings, while the solid line (refined) refers to results with the refined speaker embeddings.}
\label{fig:dvector-speaker-num}
\end{figure}

We first evaluate the refined embedding for speaker number detection. We search for the optimal threshold for this task through simulated conversations on the validation set. The threshold which achieves the lowest mean speaker number detection error across all simulated sessions is chosen for evaluation on testing set. 
As shown in Figure~\ref{fig:dvector-speaker-num}, the minimal error achieved by the refined embeddings is lower than the original embeddings with a threshold of $3.0$ and $2.0$, respectively. The curved slope of the refined embeddings after the optimal threshold is lower than the original embeddings, which indicates that speaker number detection with the refined embedding is more robust.

With the optimal threshold tuned on the validation set we evaluate the performance of speaker number detection on two test sets: fake conversations simulated from the testing set of VoxCeleb2 and an in-house meeting dataset, which are in-domain and out-of-domain datasets respectively. As shown in Table~\ref{tbl:d-vector-results}, the performance with refined embeddings outperforms the original embeddings on both datasets with both the eigengap-based method~\cite{wang2018speaker} and the threshold-based method.

Intriguingly, while the threshold-based methods with original and refined embeddings outperform the eigengap-based methods on the in-domain dataset (VoxCeleb2 Test), the performance of the threshold-based method with original embeddings degrades on the out-of-domain (in-house meeting) dataset. This is likely because original embeddings are less robust with respect to the threshold, while  
the refined embeddings are robust even with a threshold tuned on a different dataset. Consequently the threshold-based method with refined embeddings significantly outperforms other approaches on both in-domain and out-of-domain datasets. 

\begin{table}[h]
\caption{Mean error of speaker number detection (first two rows) and DER (last row) for the eigengap-based (\textit{gap}) method and the threshold-based method (\textit{thred}) using original (O) and refined (R) embeddings, respectively. The last line shows DER improvement relative to O-\textit{gap}. }
\label{tbl:d-vector-results}
\centering
\begin{tabular}{ccccc}
\specialrule{.1em}{.05em}{.05em} 
Dataset & O-\textit{gap} & O-\textit{thred} & R-\textit{gap} & R-\textit{thred} \\ \hline
VoxCeleb2 Test & 4.77 & 1.56 & 2.64 & \textbf{1.03} \\ 
In-house meetings      & 3.23 & 4.39 & 2.86 & \textbf{1.20} \\ \hline
In-house meetings  & \multirow{2}{*}{0\%} & \multirow{2}{*}{-116.7\%} & \multirow{2}{*}{16.3\%} & \multirow{2}{*}{\textbf{73.7\%}} \\
\footnotesize{(Rel. DER reduction)} & & & & \\
\specialrule{.1em}{.05em}{.05em} 
\end{tabular}
\end{table}

\subsection{Speaker diarization results}
We compare the performance of spectral clustering on speaker diarization using both original and refined embeddings.
The results in Table~\ref{tbl:d-vector-results} are consistent with the results on speaker number detection. Here again using the threshold-based model using the original embeddings actually performs worse than the eigengap-based method for the out-of-domain dataset. However the model using refined embeddings and the threshold-based method for speaker number detection achieves a 73.7\% relative DER reduction compared to the model with the original embeddings. This indicates that our system should be generalizable to real world applications.

\section{Conclusion}
In this paper, we present the first use of GNNs for speaker diarization. A graph of speech segments is built for each meeting session and a GNN model is trained to refine the speaker embedding utilizing the structural information in the graph. Our model achieves the state-of-the-art performance on a public dataset, while experiments on both simulated and real meetings show that spectral clustering based on the refined embeddings can achieve much better performance than original embeddings in terms of speaker number detection and clustering.

\section{Acknowledgement}
The authors would like to thank Tianyan Zhou and Yong Zhao for providing the d-vector extraction model, and Chun-Hao Chang and Zhuo Chen for helpful discussions. Rudzicz is a CIFAR Chair in AI.


\section{References}
\bibliographystyle{IEEEbib}
{\footnotesize \bibliography{refs}}

\begin{thebibliography}{10}

\bibitem{dehak2010front}
Najim Dehak, Patrick~J Kenny, R{\'e}da Dehak, Pierre Dumouchel, and Pierre
  Ouellet,
\newblock ``Front-end factor analysis for speaker verification,''
\newblock {\em IEEE Transactions on Audio, Speech, and Language Processing},
  vol. 19, no. 4, pp. 788--798, 2010.

\bibitem{snyder2018x}
David Snyder, Daniel Garcia-Romero, Gregory Sell, Daniel Povey, and Sanjeev
  Khudanpur,
\newblock ``X-vectors: Robust dnn embeddings for speaker recognition,''
\newblock in {\em 2018 IEEE International Conference on Acoustics, Speech and
  Signal Processing (ICASSP)}. IEEE, 2018, pp. 5329--5333.

\bibitem{variani2014deep}
Ehsan Variani, Xin Lei, Erik McDermott, Ignacio~Lopez Moreno, and Javier
  Gonzalez-Dominguez,
\newblock ``Deep neural networks for small footprint text-dependent speaker
  verification,''
\newblock in {\em 2014 IEEE International Conference on Acoustics, Speech and
  Signal Processing (ICASSP)}. IEEE, 2014, pp. 4052--4056.

\bibitem{bredin2017tristounet}
Herv{\'e} Bredin,
\newblock ``Tristounet: triplet loss for speaker turn embedding,''
\newblock in {\em 2017 IEEE international conference on acoustics, speech and
  signal processing (ICASSP)}. IEEE, 2017, pp. 5430--5434.

\bibitem{wan2018generalized}
Li~Wan, Quan Wang, Alan Papir, and Ignacio~Lopez Moreno,
\newblock ``Generalized end-to-end loss for speaker verification,''
\newblock in {\em 2018 IEEE International Conference on Acoustics, Speech and
  Signal Processing (ICASSP)}. IEEE, 2018, pp. 4879--4883.

\bibitem{wang2019centroid}
Jixuan Wang, Kuan-Chieh Wang, Marc~T Law, Frank Rudzicz, and Michael Brudno,
\newblock ``Centroid-based deep metric learning for speaker recognition,''
\newblock in {\em ICASSP 2019-2019 IEEE International Conference on Acoustics,
  Speech and Signal Processing (ICASSP)}. IEEE, 2019, pp. 3652--3656.

\bibitem{zhou2019cnn}
Tianyan Zhou, Yong Zhao, Jinyu Li, Yifan Gong, and Jian Wu,
\newblock ``Cnn with phonetic attention for text-independent speaker
  verification,''
\newblock in {\em Proc. IEEE Workshop on Automatic Speech Recognition and
  Understanding}, 2019.

\bibitem{meignier2010lium}
Sylvain Meignier and Teva Merlin,
\newblock ``Lium spkdiarization: an open source toolkit for diarization,''
\newblock 2010.

\bibitem{shum2013unsupervised}
Stephen~H Shum, Najim Dehak, R{\'e}da Dehak, and James~R Glass,
\newblock ``Unsupervised methods for speaker diarization: An integrated and
  iterative approach,''
\newblock {\em IEEE Transactions on Audio, Speech, and Language Processing},
  vol. 21, no. 10, pp. 2015--2028, 2013.

\bibitem{sell2014speaker}
Gregory Sell and Daniel Garcia-Romero,
\newblock ``Speaker diarization with plda i-vector scoring and unsupervised
  calibration,''
\newblock in {\em 2014 IEEE Spoken Language Technology Workshop (SLT)}. IEEE,
  2014, pp. 413--417.

\bibitem{garcia2017speaker}
Daniel Garcia-Romero, David Snyder, Gregory Sell, Daniel Povey, and Alan
  McCree,
\newblock ``Speaker diarization using deep neural network embeddings,''
\newblock in {\em 2017 IEEE International Conference on Acoustics, Speech and
  Signal Processing (ICASSP)}. IEEE, 2017, pp. 4930--4934.

\bibitem{wang2018speaker}
Quan Wang, Carlton Downey, Li~Wan, Philip~Andrew Mansfield, and Ignacio~Lopz
  Moreno,
\newblock ``Speaker diarization with lstm,''
\newblock in {\em 2018 IEEE International Conference on Acoustics, Speech and
  Signal Processing (ICASSP)}. IEEE, 2018, pp. 5239--5243.

\bibitem{zhang2019fully}
Aonan Zhang, Quan Wang, Zhenyao Zhu, John Paisley, and Chong Wang,
\newblock ``Fully supervised speaker diarization,''
\newblock in {\em ICASSP 2019-2019 IEEE International Conference on Acoustics,
  Speech and Signal Processing (ICASSP)}. IEEE, 2019, pp. 6301--6305.

\bibitem{lloyd1982least}
Stuart Lloyd,
\newblock ``Least squares quantization in pcm,''
\newblock {\em IEEE transactions on information theory}, vol. 28, no. 2, pp.
  129--137, 1982.

\bibitem{rokach2005clustering}
Lior Rokach and Oded Maimon,
\newblock ``Clustering methods,''
\newblock in {\em Data mining and knowledge discovery handbook}, pp. 321--352.
  Springer, 2005.

\bibitem{von2007tutorial}
Ulrike Von~Luxburg,
\newblock ``A tutorial on spectral clustering,''
\newblock {\em Statistics and computing}, vol. 17, no. 4, pp. 395--416, 2007.

\bibitem{frey2007clustering}
Brendan~J Frey and Delbert Dueck,
\newblock ``Clustering by passing messages between data points,''
\newblock {\em science}, vol. 315, no. 5814, pp. 972--976, 2007.

\bibitem{fujita2019end}
Yusuke Fujita, Naoyuki Kanda, Shota Horiguchi, Kenji Nagamatsu, and Shinji
  Watanabe,
\newblock ``End-to-end neural speaker diarization with permutation-free
  objectives,''
\newblock {\em Proc. Interspeech 2019}, pp. 4300--4304, 2019.

\bibitem{kipf2016variational}
Thomas~N Kipf and Max Welling,
\newblock ``Variational graph auto-encoders,''
\newblock {\em arXiv preprint arXiv:1611.07308}, 2016.

\bibitem{zhang2018link}
Muhan Zhang and Yixin Chen,
\newblock ``Link prediction based on graph neural networks,''
\newblock in {\em Advances in Neural Information Processing Systems}, 2018, pp.
  5165--5175.

\bibitem{nickel2015review}
Maximilian Nickel, Kevin Murphy, Volker Tresp, and Evgeniy Gabrilovich,
\newblock ``A review of relational machine learning for knowledge graphs,''
\newblock {\em Proceedings of the IEEE}, vol. 104, no. 1, pp. 11--33, 2015.

\bibitem{wu2019comprehensive}
Zonghan Wu, Shirui Pan, Fengwen Chen, Guodong Long, Chengqi Zhang, and Philip~S
  Yu,
\newblock ``A comprehensive survey on graph neural networks,''
\newblock {\em arXiv preprint arXiv:1901.00596}, 2019.

\bibitem{cai2018comprehensive}
Hongyun Cai, Vincent~W Zheng, and Kevin Chen-Chuan Chang,
\newblock ``A comprehensive survey of graph embedding: Problems, techniques,
  and applications,''
\newblock {\em IEEE Transactions on Knowledge and Data Engineering}, vol. 30,
  no. 9, pp. 1616--1637, 2018.

\bibitem{kipf2016semi}
Thomas~N. Kipf and Max Welling,
\newblock ``Semi-supervised classification with graph convolutional networks,''
\newblock in {\em 5th International Conference on Learning Representations,
  {ICLR} 2017, Toulon, France, April 24-26, 2017, Conference Track
  Proceedings}. 2017, OpenReview.net.

\bibitem{gilmer2017neural}
Justin Gilmer, Samuel~S Schoenholz, Patrick~F Riley, Oriol Vinyals, and
  George~E Dahl,
\newblock ``Neural message passing for quantum chemistry,''
\newblock in {\em Proceedings of the 34th International Conference on Machine
  Learning-Volume 70}. JMLR. org, 2017, pp. 1263--1272.

\bibitem{lin2019lstm}
Qingjian Lin, Ruiqing Yin, Ming Li, Herv{\'e} Bredin, and Claude Barras,
\newblock ``{LSTM based Similarity Measurement with Spectral Clustering for
  Speaker Diarization},''
\newblock 2019, {HAL CCSD}.

\bibitem{ustinova2016learning}
Evgeniya Ustinova and Victor Lempitsky,
\newblock ``Learning deep embeddings with histogram loss,''
\newblock in {\em Advances in Neural Information Processing Systems}, 2016, pp.
  4170--4178.

\bibitem{recht2010guaranteed}
Benjamin Recht, Maryam Fazel, and Pablo~A Parrilo,
\newblock ``Guaranteed minimum-rank solutions of linear matrix equations via
  nuclear norm minimization,''
\newblock {\em SIAM review}, vol. 52, no. 3, pp. 471--501, 2010.

\bibitem{callhome_receipe}
``\textit{callhome\_diarization/v2} receipe of {Kaldi},''
  \url{https://github.com/kaldi-asr/kaldi/tree/master/egs/callhome\_diarization/v2},
\newblock Accessed: 2019-10-15.

\bibitem{xvector_model}
``{SRE16 Xvector Model},'' \url{http://kaldi-asr.org/models/m6},
\newblock Accessed: 2019-10-15.

\bibitem{clevert2015fast}
Djork{-}Arn{\'{e}} Clevert, Thomas Unterthiner, and Sepp Hochreiter,
\newblock ``Fast and accurate deep network learning by exponential linear units
  (elus),''
\newblock in {\em 4th International Conference on Learning Representations,
  {ICLR}}, 2016.

\bibitem{fey2019fast}
Matthias Fey and Jan~Eric Lenssen,
\newblock ``Fast graph representation learning with pytorch geometric,''
\newblock {\em arXiv preprint arXiv:1903.02428}, 2019.

\bibitem{7178881}
G.~{Sell} and D.~{Garcia-Romero},
\newblock ``Diarization resegmentation in the factor analysis subspace,''
\newblock in {\em 2015 IEEE International Conference on Acoustics, Speech and
  Signal Processing (ICASSP)}, April 2015, pp. 4794--4798.

\bibitem{7953094}
D.~{Garcia-Romero}, D.~{Snyder}, G.~{Sell}, D.~{Povey}, and A.~{McCree},
\newblock ``Speaker diarization using deep neural network embeddings,''
\newblock in {\em 2017 IEEE International Conference on Acoustics, Speech and
  Signal Processing (ICASSP)}, March 2017, pp. 4930--4934.

\bibitem{chung2018voxceleb2}
Joon~Son Chung, Arsha Nagrani, and Andrew Zisserman,
\newblock ``Voxceleb2: Deep speaker recognition,''
\newblock in {\em Proc. Interspeech 2018}, 2018, pp. 1086--1090.

\end{thebibliography}

\end{document}